\documentclass[12pt]{article}

\usepackage{graphics}
\usepackage{epsfig}

\begin{document}

\title{Trust in the CODA model:\\ Opinion Dynamics and the reliability of other agents
}
\author{Andr\'e C. R. Martins\\
NISC - EACH\\ Universidade de S\~ao Paulo, Brazil
}
\maketitle

\begin{abstract}
A model for the joint evolution of opinions and how much the agents trust each other is presented. The model is built using the framework of the Continuous Opinions and Discrete Actions (CODA) model. Instead of a fixed probability that the other agents will decide in the favor of the best choice, each agent considers that other agents might be one one of two types: trustworthy or useless. Trustworthy agents are considered more likely to be right than wrong, while the opposite holds for useless ones. Together with the opinion about the discussed issue, each agent also updates that probability for each one of the other agents it interacts withe probability each one it interacts with is of one type or the other. The dynamics of opinions and the evolution of the trust between the agents are studied. Clear evidences of the existence of two phases, one where strong polarization is observed and the other where a clear division is permanent and reinforced are observed. The transition seems signs of being a first-order transition, with a location dependent on both the parameters of the model and the initial conditions. This happens despite the fact that the trust network evolves much slower than the opinion on the central issue.

\end{abstract}

\section{Introduction}

Opinion Dynamics models \cite{castellanoetal07} are usually built on the idea that each agent will influence others, either all of them, if one assumes a complete graph of connections, or a set of all agents that compose its neighborhood. In the most common case, agents somehow change their opinions towards the opinion of those neighbors, either by a simple imitative process in discrete models \cite{galametal82,galammoscovici91,sznajd00,stauffer03a} or moving the values of their opinion in the direction of the value of the neighbor opinion, in the continuous models \cite{deffuantetal00,hegselmannkrause02}. A mixed version exists in the Continuous Opinions and Discrete Action (CODA) model \cite{martins08a,martins12b}, where the continuous values are not observed and only a discrete choice is known by the neighbors. CODA model variations have been shown to be equivalent to the continuous models \cite{martins08c} as well as a general case to the discrete models in the literature \cite{martins12a}.

However, it is not true that people always tend to copy those they observe. Depending on their own characteristics and those of the person one is interacting with, observation might either have no consequences or make the person who observes want to avoid the observed characteristic \cite{sobkowiczsobkowicz10a}. This idea has been implemented in different ways in different models. No influence has been coded as a threshold in the continuous Bounded Confidence models and as inflexible agents who don't change their opinions in discrete \cite{galam05,galamjacobs07,galam11a} and mixed \cite{martinsgalam13a} models. The negative influence has also been studied by the introduction of ``contrarians'' \cite{galam04,martinskuba09a} who change their opinions in the opposite direction of the influence on them. In more general terms, the problem of trust in Opinion Dynamics models has been explored for a number of different models \cite{boeroetal10a,carlettietal10a,sietal10a,sietal10b,richterspeixoto11a,fanetal21a}. In most cases mentioned here, the trust between agents is not subject to the dynamics of the models, with exception of one case \cite{carlettietal10a}.

It is clear, however, that, as our opinion about a certain subject changes, so does our opinion about the person who influenced us. And, in reverse, it is also true that we interpret the same information differently depending on its source \cite{kahanetal11}. Here, we will present a generalization of the CODA model where the agents change their opinions about the issue and also about the reliability of the other agents.

\section{Trust and Likelihood}

The original CODA model \cite{martins08c,martins08b} was obtained by assuming that, in a situation where there were two possible choices (or actions), each agent considers there is a fixed probability $\alpha > 0.5$ that each one of its neighbors will have chosen the best alternative. Let the two choices be $A$ and $B$ and let $p_i (t)$ be the probability agent $i$ assigns at time $t$ to the probability that $A$ is the best choice. CODA adopts a fixed likelihood $\alpha \equiv P(OA_j|A)$, representing the chance that, if $A$ is indeed the best choice, when observing agent $j$, $i$ will observe $j$ chose $A$, indicated here by $OA_j$. Assuming that the problem is symmetrical in relation to both choices, that is,
\[
P(OA_j|A) = P(OB_j|B),
\]
a simple use of Bayes theorem will show how $p_i(t)$ is altered. By using log-odd function $\nu \equiv \ln(\frac{p}{1-p})$ (where the agent index and time dependence were omitted for brevity), a simple additive model is obtained. This can be trivially normalized to $\nu^*$ so that when $A$ is observed, the agent adds $+1$ to $\nu^*$, and when $B$ is observed, $-1$ is added. In this case, the choice is defined simply by the sign of $\nu^*$, with positive signs indicating $A$ is chosen. Defined in this way, the model becomes actually independent of $\alpha$, except for translating the number of steps away from changing opinions back into a probability. For the non-symmetrical case, the likelihoods for choosing $A$ or $B$ could be different and, in this case, the negative and positive steps will no longer have the same size. While, in principle, one could use a normalization where one of the steps is changed to a size of one, there is no strong reason to do that, as the final model will no longer be as simple as the symmetric case. 

Contrarians were previously introduced in the symmetrical case simply by assigning a percentage of agents who actually reacted with the opposite sign from the original model \cite{martinskuba09a}. This is actually equivalent to a supposition that the contrarian neighbors are more likely to be wrong than right, that is, they actually have a likelihood of being right given by
\begin{equation}
\mu = 1 - \alpha < 0.5.
\end{equation}
Of course, the assumption that $\mu$ and $\alpha$ add to one is a simplifying one. It actually makes no difference in the dynamics of choices, since, for each agent, in that model, one could define the renormalized number of steps according to the agent own likelihoods.
This was due to the fact that each agent treated all other agents it interacted with in the same way, contrarians, in the model, simply believed their neighbors were more likely to be wrong than right.

However, if one wants to introduce the possibility that people trust some people more than others, it makes sense to introduce a trust matrix $\tau_{ij}$ that represents how much agent $i$ thinks agent $j$ is likely to be more reliable than useless. That is, according to agent $i$ there is a probability $\tau_{ij}$ that the choices of agent $j$ have a probability $\alpha >0.5$ of being right (that is that agent $j$ is trustworthy ($T$)) and a probability $1-\tau_{ij}$ that those choices have a chance $\mu <0.5$ of being the best ones ($j$ is useless $U$). Each agent believes there are two types of agents, $T$ who are likely to make the right choices, and useless $U$, more likely to choose wrong. In both cases, the agents realize that there is a probability that each of its neighbors will not act according to its character, that is, trustworthy agents have a chance to be wrong ($1-\alpha$) and useless agents have a chance to pick the best choice randomly also ($1-\mu$).

Obtaining the update rule in this case is a simple application of the Bayes Theorem, as per the framework in the CODA model \cite{martins12b}. Each agent $i$ believes at time $t$ the chance that $A$ is the best choice is given by $p_i (t)$. At that time, it observes an agent $j$, whom $i$ believes has a probability $\tau_{ij}$ of being of the $T$ type, meaning $j$ would chose $A$ with probability $\alpha$ and $B$ with probability $1-\alpha$. If agent $j$ is a type $U$, which happens with probability $1-\tau_{ij}$, it would, instead, choose $A$ with probability $\mu$ and $B$ with probability $1-\mu$.

Assuming that $OA_j$ is observed, that is, that $j$ is observed to prefer $A$, agent $i$ will update its probability $p_i (t)$ to $p_i(t+1)$ obtained by applying Bayes Theorem and given by
\begin{equation}\label{eq:pupdatea}
p_i(t+1) = 
\frac{p\left[ \tau_{ij} \alpha + (1-\tau_{ij})\mu \right]}{p\left[ \tau_{ij}\alpha + (1-\tau_{ij})\mu \right] + (1-p)
\left[ \tau_{ij}(1-\alpha) + (1-\tau_{ij})(1-\mu ) \right]},
\end{equation}
where $p$ is written in the place of $p_i(t)$ for brevity sake. One interesting effect here is that Bayes Theorem also applies to $\tau_{ij}$ and the agent $i$ also updates its opinion about how likely $j$ is to be trustworthy. We have a new $\tau_{ij} (t+1)$ given by
\begin{equation}\label{eq:tauupdatea}
\tau_{ij}(t+1) =  
\frac{\tau_{ij}\left[ p\alpha + (1-p)(1-\alpha ) \right]}{\tau_{ij}\left[ p\alpha + (1-p)(1-\alpha ) \right] + (1-\tau_{ij})
\left[ p\mu + (1-p)(1-\mu ) \right]} .
\end{equation}

It makes sense to ask what happens if one adopts the same transformation to log-odds. By calculating $p/(1-p)$, one does get rid of the numerator and taking the logarithm does lead us to an additive model. That is, for $\nu_i (t)= \ln (p/(1-p))$, we have
\[
\nu_i (t+1)= \nu_i (t)+ \ln \left[
\frac{\tau_{ij} \alpha + (1-\tau_{ij})\mu}{\tau_{ij}(1-\alpha) + (1-\tau_{ij})(1-\mu )}
\right] .
\]
If one takes $\tau_{ij} = 1$, that is, certainty that all neighbors are not trustworthy, the equation above has a simple constant additive part as second term, the constant size of the step. However, in the general case, the term will change at each interaction since it depends on $\tau_{ij}$. Similarly, the additive term equation for $\theta \equiv \ln (\tau / (1-\tau))$ depends on $p$ and will change with the dynamics. This makes it simpler to work directly with the probabilities $p$ and $\tau$, instead of using log-odds. 

In the CODA original model, using $p$ as the main variable was not a good idea also because it approached 0 or 1 too fast, so that computational limitations on the representation of real numbers became a real problematic issue. As a matter of fact, it was usual to obtain values like $1-10^{-300}$, that, in any computer, is indistinguishable from 1.0. The same was not true for $\nu$, meaning it was a much better variable to use. As we will see bellow in the simulation results, when introducing the possibility the neighbor could be useless, the probabilities do not move to such problematic zones  and, therefore, we will not use log-odds from now on, with no trouble other than the lack of simplification of the final model.

Finally, if, instead of $OA_j$, we have that agent $j$ preference is $B$, with $OB_j$ being observed, we have, instead of Equations \ref{eq:pupdatea} and \ref{eq:tauupdatea}, the following update rules:
\begin{equation}\label{eq:pupdateb}
p_i(t+1) = 
\frac{p\left[ \tau_{ij} (1-\alpha) + (1-\tau_{ij})(1-\mu) \right]}{p\left[ \tau_{ij}(1-\alpha) + (1-\tau_{ij})(1-\mu) \right] + (1-p)
\left[ \tau_{ij}\alpha + (1-\tau_{ij})\mu \right]},
\end{equation}
and
\begin{equation}\label{eq:tauupdateb}
\tau_{ij}(t+1) =  
\frac{\tau_{ij}\left[ p(1-\alpha) + (1-p)\alpha \right]}{\tau_{ij}\left[ p(1-\alpha ) + (1-p)\alpha  \right] + (1-\tau_{ij})
\left[ p(1-\mu ) + (1-p)\mu  \right]} .
\end{equation}

\begin{figure}
\centering
\begin{tabular}{cc}
\epsfig{file=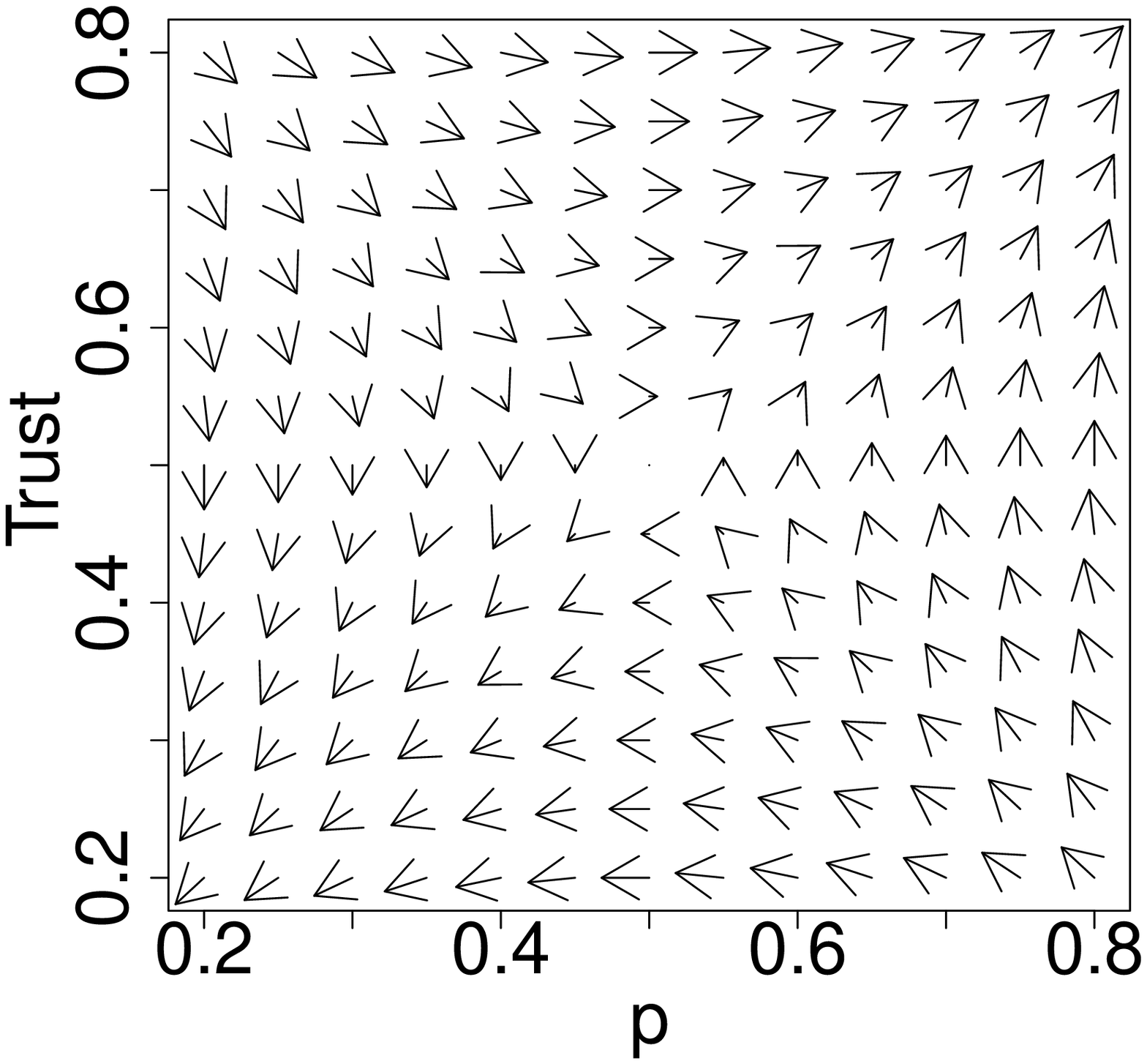,width=0.45\linewidth,clip=} & 
\epsfig{file=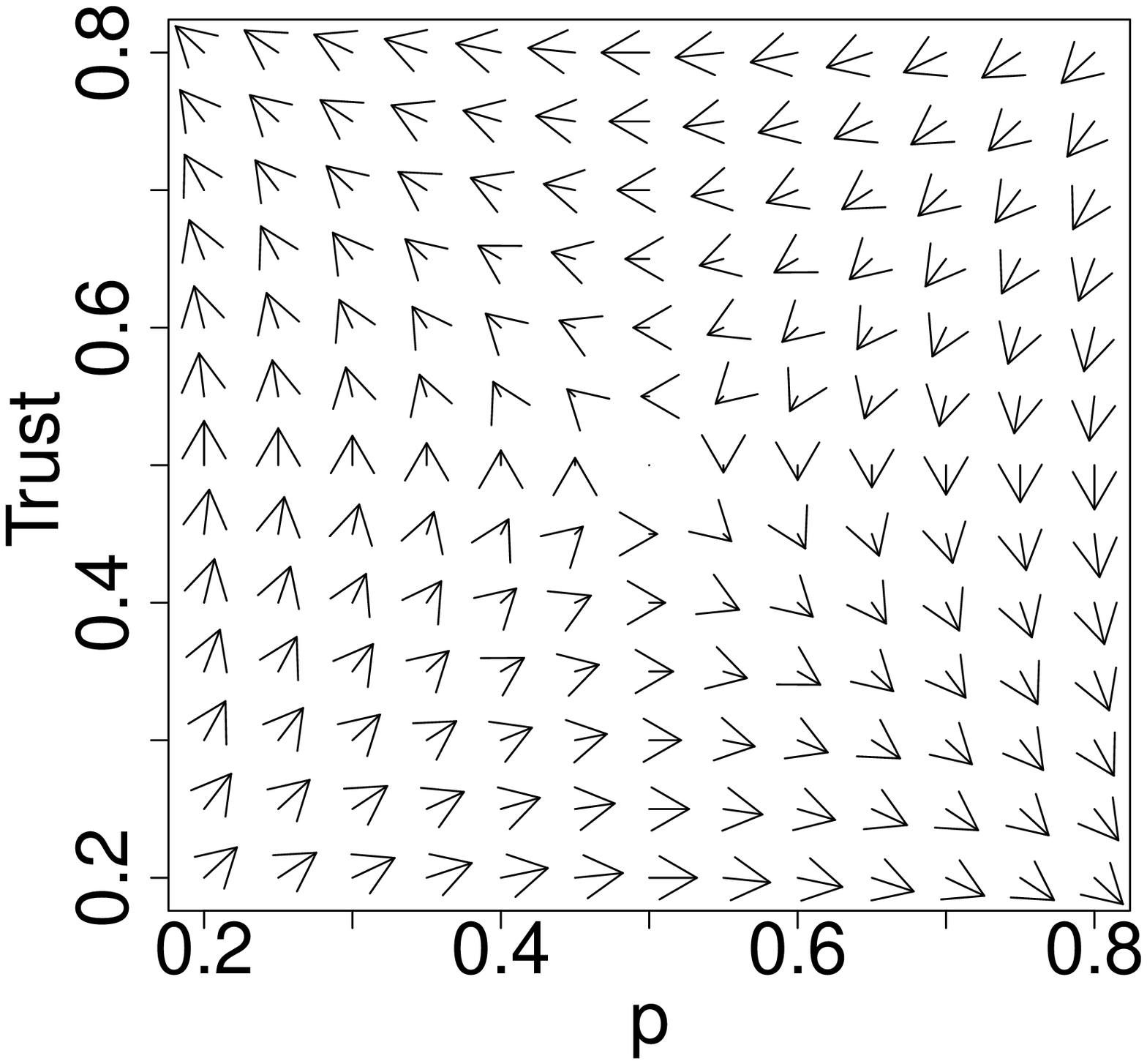,width=0.45\linewidth,clip=}
\end{tabular}
\caption{Vector field showing the direction of the effects of one update for $\alpha=0.6$ and $\mu=0.4$ depending on the different initial opinions of agent $i$. {\it Left Panel}: Agent $j$ prefers $A$.  {\it Right Panel}: Agent $j$ prefers $B$.}\label{fig:vecfield0604}
\end{figure}

\begin{figure}
\centering
\begin{tabular}{cc}
\epsfig{file=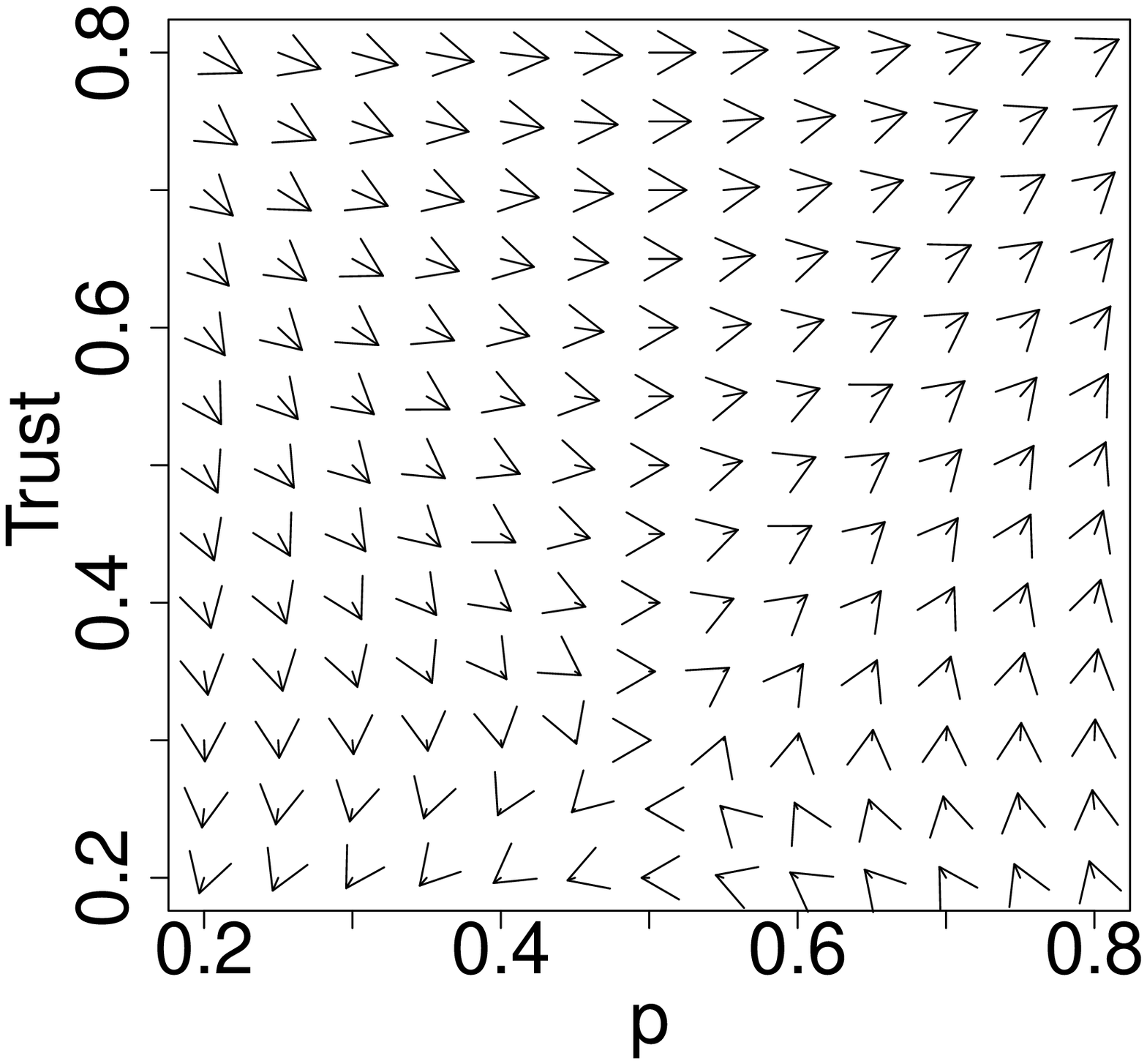,width=0.45\linewidth,clip=} & 
\epsfig{file=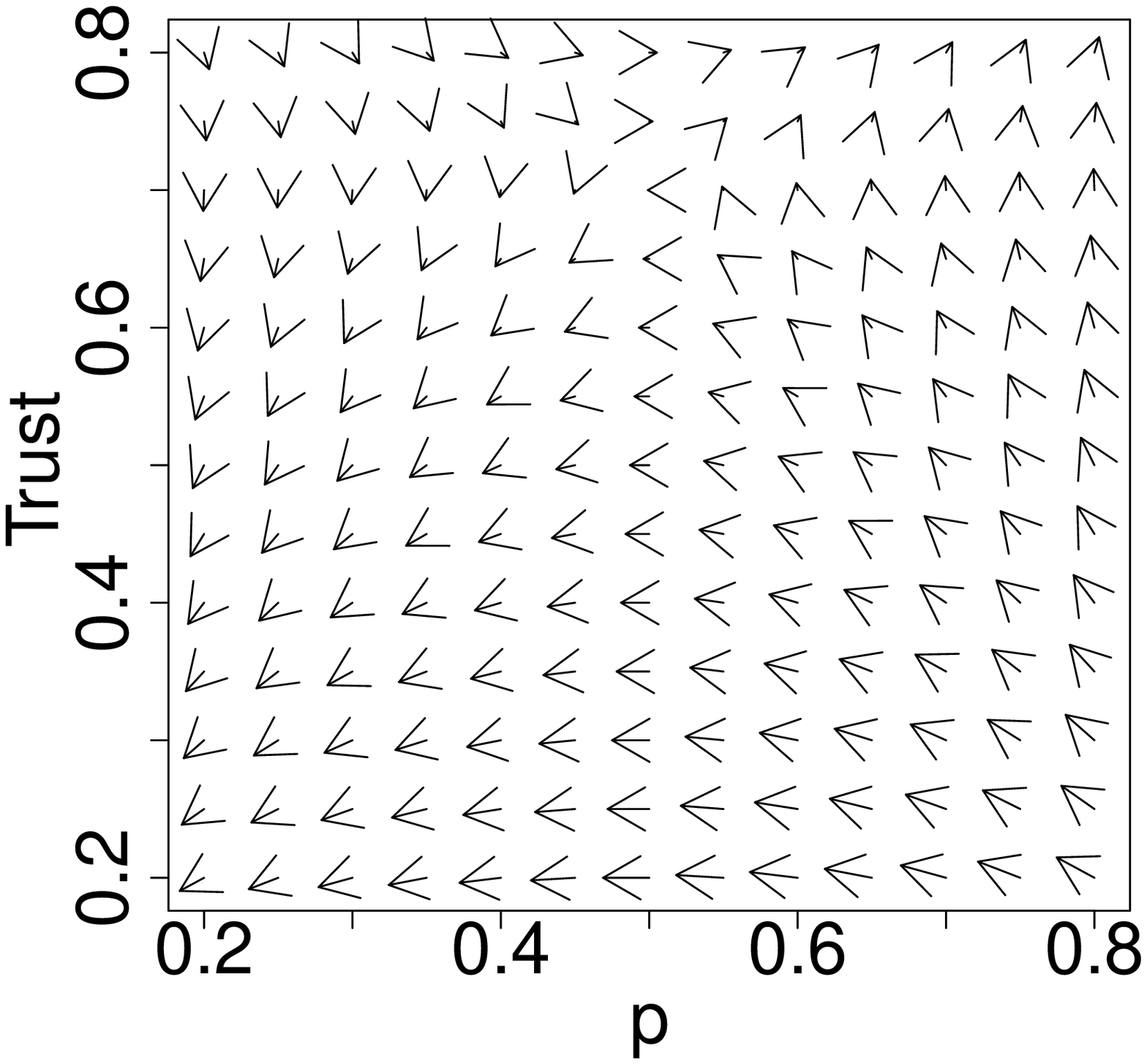,width=0.45\linewidth,clip=}
\end{tabular}
\caption{Vector field showing the direction of the effects of one update depending on the different initial opinions of agent $i$. In every case, agent $j$ prefers $A$. {\it Left Panel}: $\alpha=0.75$ and $\mu=0.4$.  {\it Right Panel}: $\alpha=0.6$ and $\mu=0.25$.}\label{fig:vecfieldja}
\end{figure}

The general effects of one single update can be seen in the Figures \ref{fig:vecfield0604} and \ref{fig:vecfieldja}. Figure \ref{fig:vecfield0604} shows the symmetrical case where $\alpha=1-\mu$, showing between the two panels, the difference in effect that one obtains by agent $j$ choosing either $A$ or $B$. As we can see, the reasonable assumption that the decisions will have opposite effects is shown to be a good one, although a numerical inspection will show they do not cancel each other exactly. We can also see that the fact that agent $j$ prefers $A$ will only make $A$ more likely for $i$, increasing $p_i$, when $i$ trusts $j$ to be more trustworthy than useless ($\tau_{ij}>0.5$). Otherwise, $i$ will move in the opposite direction to that of $j$ opinion, making it a local contrarian.

Figure \ref{fig:vecfieldja} shows us two non-symmetrical cases ($\alpha \neq 1-\mu$). As we can see, the effect of the type of agent with the likelhood ($\alpha$ or $\mu$) further from 0.5 is stronger. This translates to the fact that, when trustworthy agents are considered to be certain more often and useless ones are closer to a coin flip, only for very low values of trust will agent $i$ move as a contrarian. The opposite happens when the useless agents are considered to choose the worst choice more often than the trustworthy ones choose the better one.

One should notice that these are just the results of one update. As the system evolves, some agents will randomly interact more often with agents choosing $A$ and others with agents choosing $B$. This means each one will be, at first, led to a different region of the $p-\tau$ plane and, therefore, it is not clear, just from looking at the arrow fields, what will happen in the system as a whole.

\section{Simulation results}

The original CODA model was, in previous works, defined on a network that dictated the neighbors that each agent could interact with. This was necessary as it was the appearance of local self-reinforcing neighborhoods that kept consensus from appearing and allowed extreme different beliefs to evolve and co-exist. A fully connected model would trivially lead to a consensus around the random size that started bigger, with all agents strongly supporting it after a while.

Here, however, we are also interested in the dynamics of the trust between the agents. In the original model, we can consider that only the pre-assigned neighbors were considered to have information worthy of obtaining. But, in the general case, that can be an effect of the dynamics. In order to explore that, complete graphs will be assumed in the simulations presented here, so that every agent will be able to interact with every other agent. The goal is to explore if consensus appears or if the agents will naturally divide themselves into groups of agents with similar opinions.

Simulations were performed using the software R \cite{Rsoftware}; the package ``fields'' \cite{furreretal10a} was used to generate the arrow graphics. In the simulations, two distinct agents were randomly drawn at each step and the first agent updated it probabilities from the observation of the second. Time $T$ is measured as the number of steps divided by the number of agents $N$, so that advancing T by one means that, on average, each agent observed one other agent once. The values of $p$ in most simulations were randomly drawn from a uniform distribution between 0.01 and 0.99, except when noted otherwise.

\begin{figure}
\centering
\begin{tabular}{ccc}
\epsfig{file=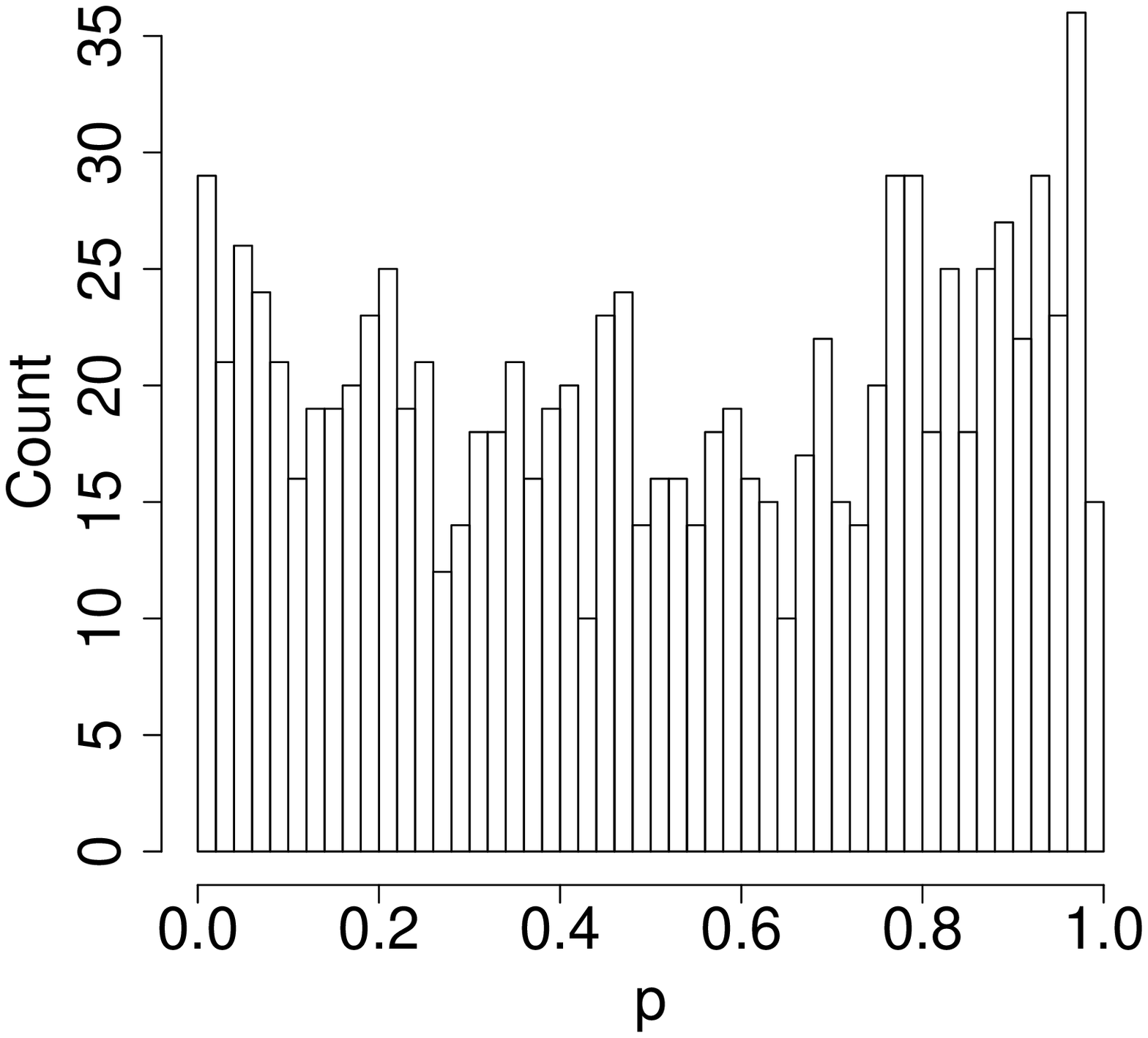,width=0.31\linewidth,clip=} & 
\epsfig{file=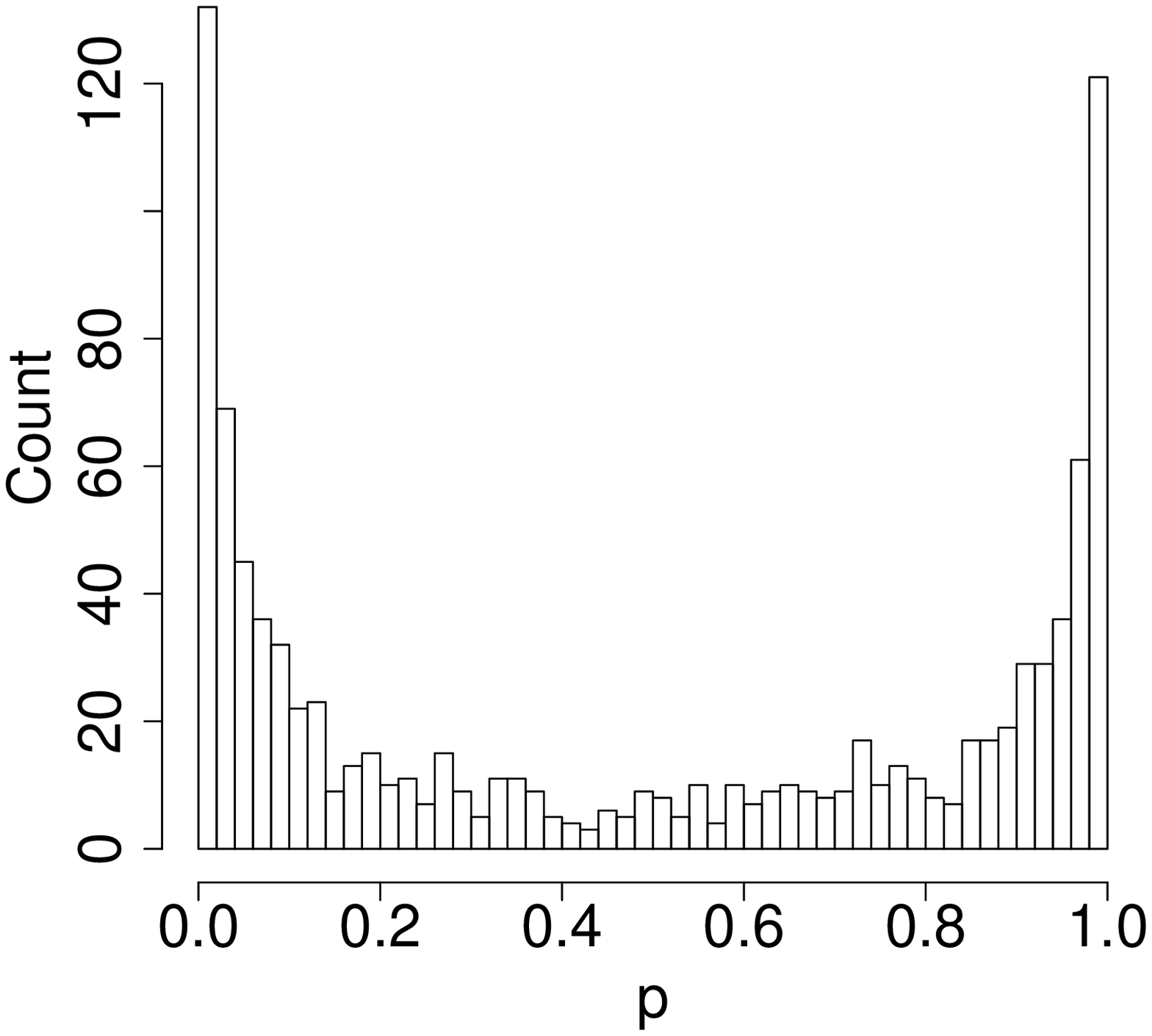,width=0.31\linewidth,clip=} & 
\epsfig{file=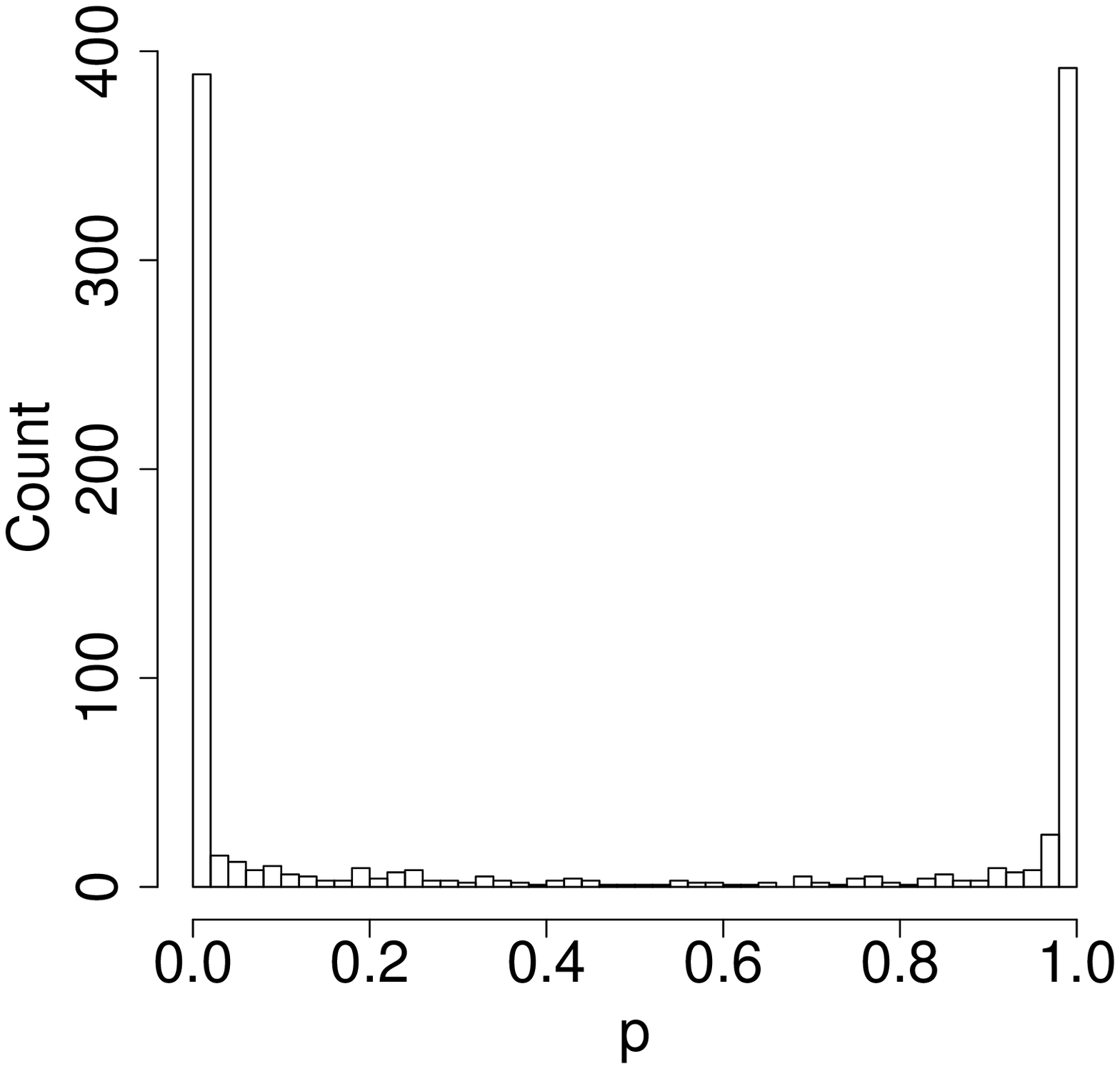,width=0.31\linewidth,clip=} 
\end{tabular}
\caption{Temporal evolution of the histograms showing the distribution of observed $p_i$ after different amounts of simulation time for $N=1,000$. In all cases, $\alpha=1-\mu=0.55$ and all agents started with $\tau_{ij}=0.5$ for all values of $i$ and $j$. {\it Left Panel}: $T=200$. {\it Central Panel}: $T=500$.  {\it Right Panel}: $T=1,000$.}\label{fig:hista55mu45}
\end{figure}

Figure \ref{fig:hista55mu45} shows the temporal evolution of one realization of a symmetric case, where $\alpha=1-\mu=0.55$ and all $N=1,000$ agents starting with $\tau_{ij}=0.5$ for all values of $i$ and $j$.  We see that, in this case, at first, even after an average of 200 interactions per agent, the probabilities are still randomly distributed over all possible values. As a comparison, without trust, the CODA model showed regions starting to stabilize with strong opinions as soon as $T=20$. Around $T=50$, most opinions were already very strong ($p$ close to 0 or 1) and the general format of the distribution of choices would remain the same from that point on \cite{martins08a}.

With the introduction of trust, we see that the opinions eventually tend to the extremes, but even at $T=500$, we still have a significant but diminishing number of agents with moderate opinions. However, those moderate opinions tend to disappear, as we can see from the case where $T=1,000$. And this is another important difference, since moderate opinions never really disappeared in the original CODA algorithm, as they survived in the interface between the extreme regions. Here, on the other hand, all agents eventually make up their minds and, as we will see, learn to mistrust those who disagree with them.

\begin{figure}
\centering
\begin{tabular}{cc}
\epsfig{file=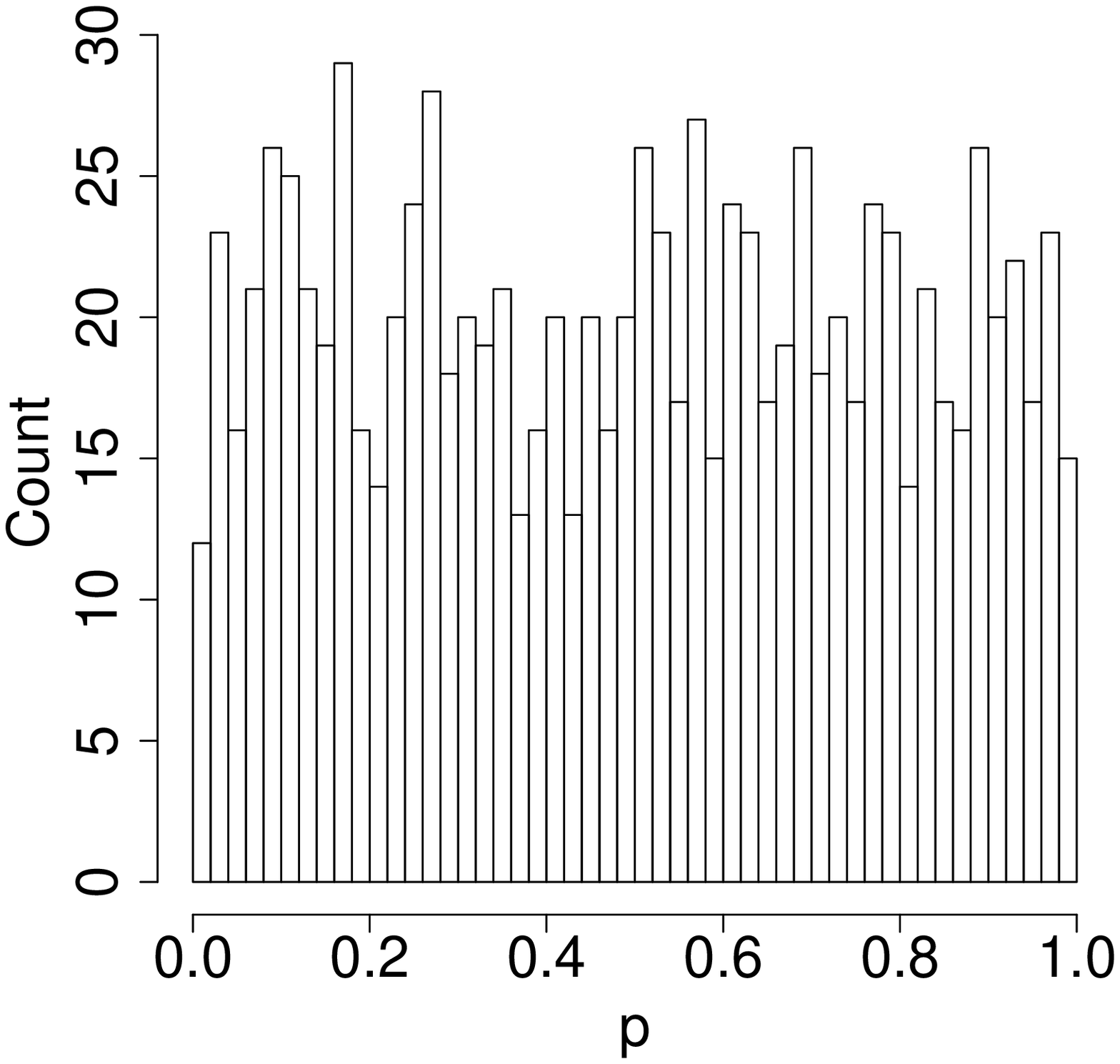,width=0.31\linewidth,clip=} & 
\epsfig{file=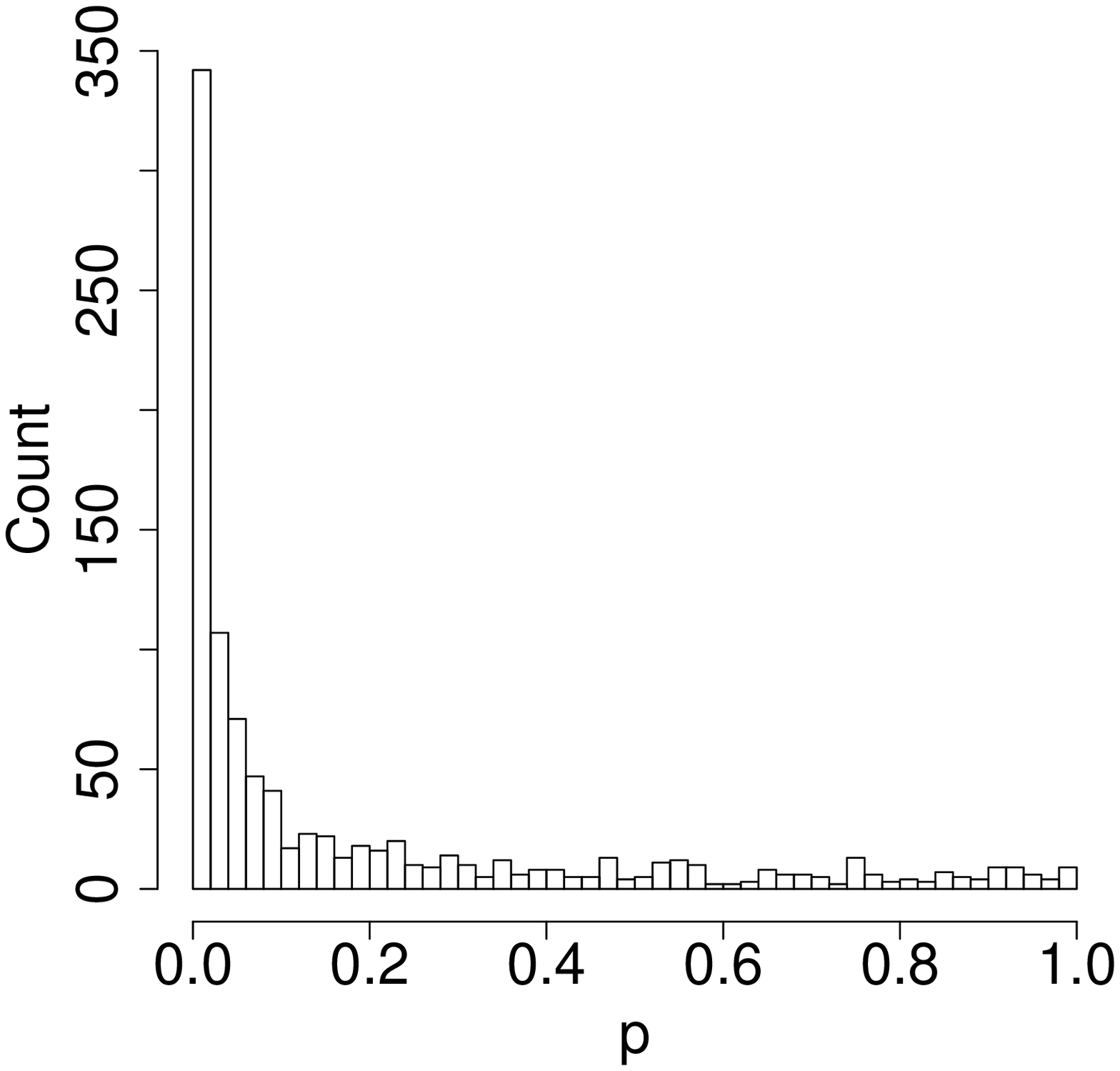,width=0.31\linewidth,clip=} 
\end{tabular}
\caption{Histograms showing the distribution of observed $p_i$ for two different cases when $T=50$. In all cases, $\alpha=1-\mu=0.55$ and all agents started with $\tau_{ij}=0.5$ for all values of $i$ and $j$. {\it Left Panel}: $\alpha=1-\mu=0.55$. {\it Right Panel}: $\alpha=0.70$ and $\mu=0.55$}\label{fig:histt50}
\end{figure}

However, unlike previous implementations, the introduction of trust allows a different behavior to appear, depending on the parameters. Figure \ref{fig:histt50} shows the distribution when $T=50$ for two different cases, $\alpha=1-\mu=0.55$ at left and $\alpha=0.70$ and $\mu=0.55$ at right. What we see is that, in the first case, as expected, too little time has passed and the system is at a similar, undecided state, as it will still be when $T=200$. However, the situation for the second, non-symmetrical case, is quite different. One peak of strong opinions is clearly formed near $p=0$, but none appeared around $p=1.0$. As a matter of fact, very few agents keep an opinion in favor of $A$ ($p>0.5$) and the system is close to reaching a consensus. Running that case a little longer indeed leads the system to a full agreement, in this case, in favor of $B$. This happens because the agents consider that the information of trustworthy agents is far more reliable than the contrarian information of the useless ones, that is $\mu$ is much closer to 0.5 than $\alpha$. This initial better evaluation of their peers makes it possible for the agents to reach an agreement before they start having strong opinions and/or mistrusting each other.

\begin{figure}
\centering
\begin{tabular}{cc}
\epsfig{file=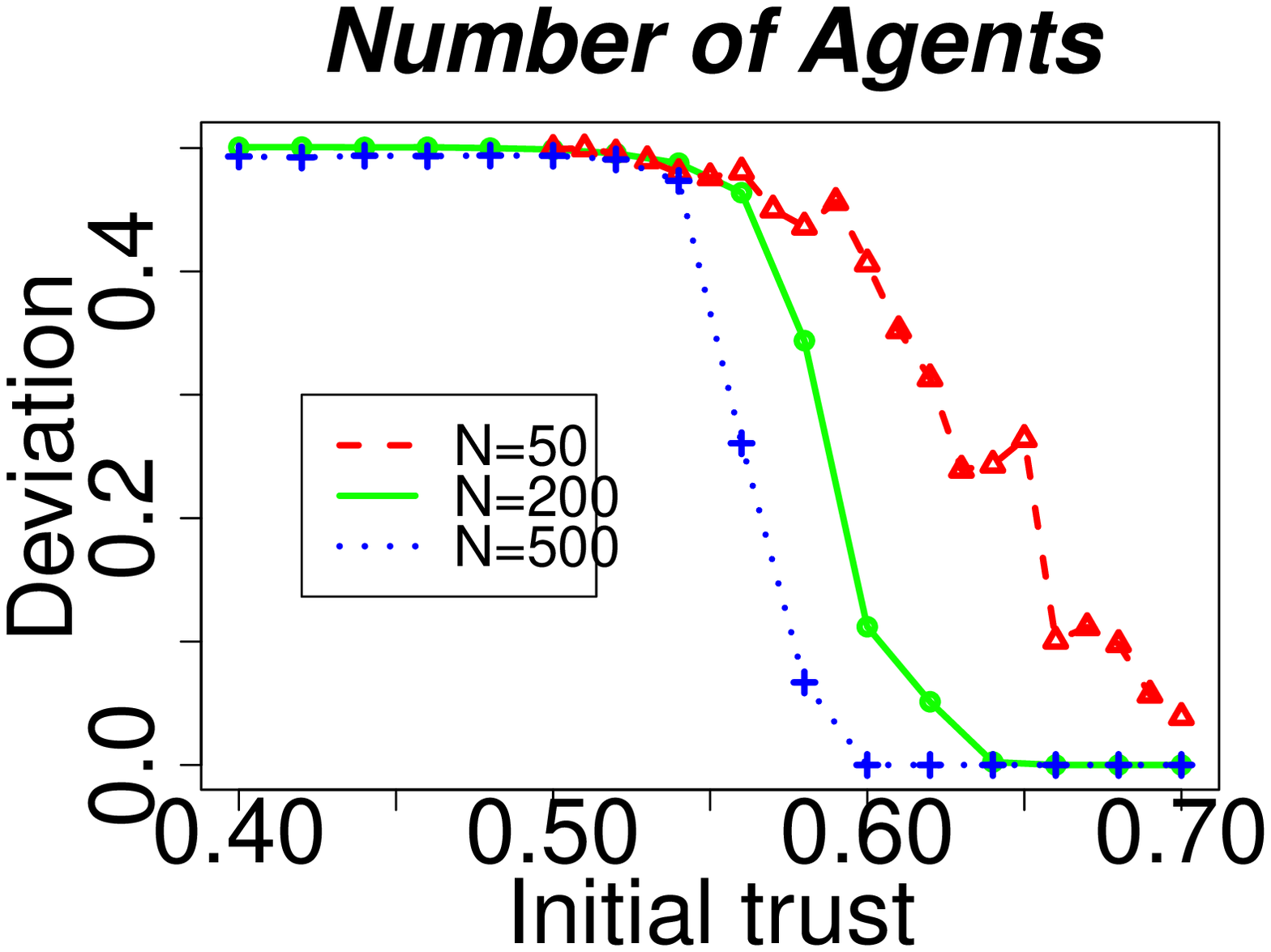,width=0.35\linewidth,clip=} & 
\epsfig{file=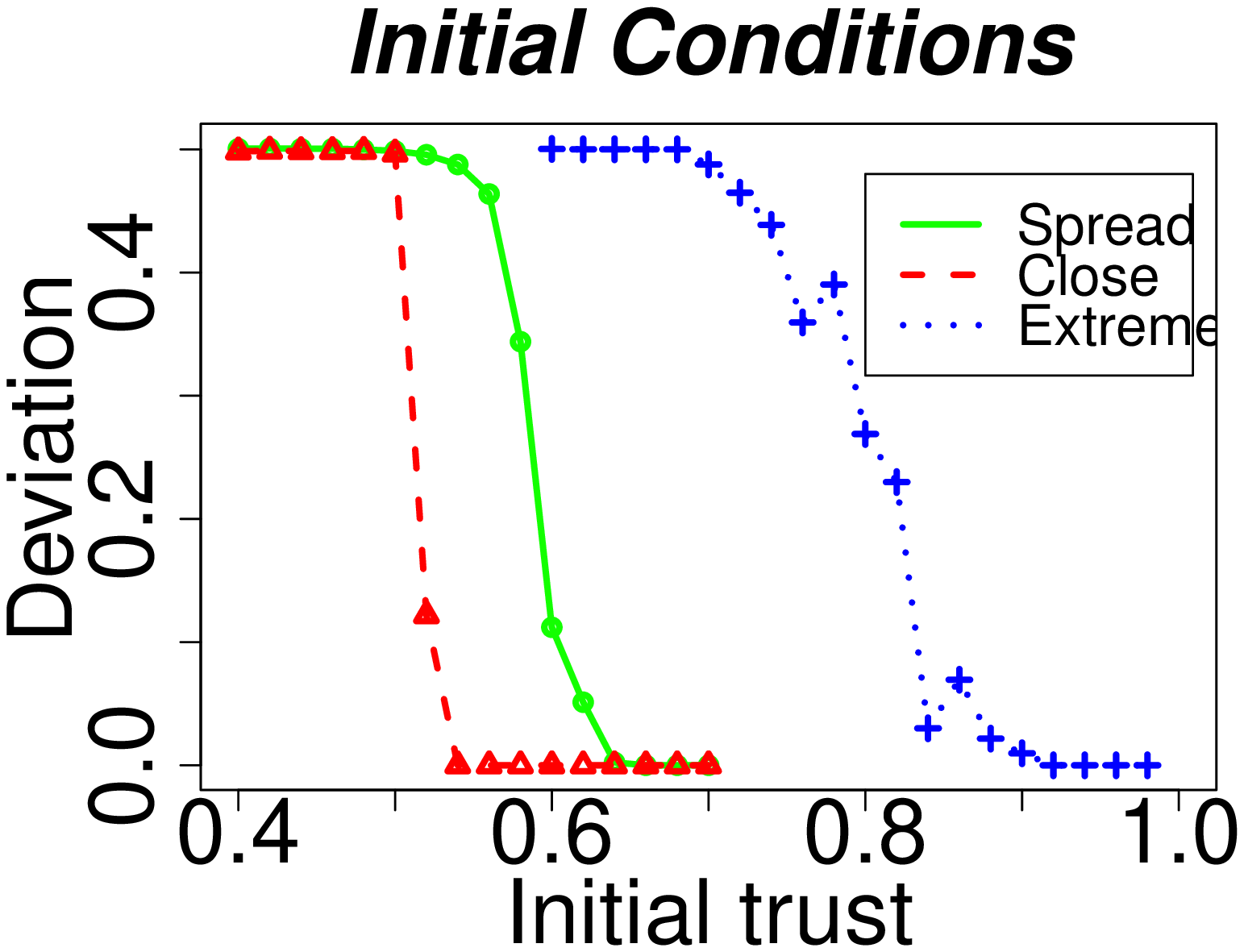,width=0.35\linewidth,clip=}
\end{tabular}
\caption{Standard deviation of $p_i$ as a function of initial trust $\tau_{ij}$. In all cases, $\alpha=1-\mu=0.55$ {\it Left Panel}: Effect of different number of agents $N$. {\it Right Panel}: Different initial conditions for $p_i$ are shown. In the spread condition, each $p_i$ was randomly drawn from a uniform distribution between 0.01 and 0.99; for the close case, between 0.4 and 0.6, while the extreme case had $p_i$ drawn uniformly in the regions between 0 and 0.01 and between 0.99 and 1.0.}\label{fig:devs}
\end{figure}

Figure \ref{fig:devs} shows the final results for different cases. All graphics depict the average over $20$ realization of the observed standard deviation of $p_i$ as a function of the initial trust $\tau_{ij}$, chosen equal for all agents at the beginning. When all agents agree, all $p_i$ are equal or very close and, therefore, the standard deviation tends to zero. The opposite extreme has half of agents with $p_i =0 $ and the other half with $p_i =1$, for a standard deviation of $0.5$. The left panel shows the same case run with different number $N$ of agents and $T=1,000$. What we see resembles a finite case associated with a first order transition between complete agreement among the agents, when the trust starts high enough, and polarization between two disagreeing groups when trust starts low. The right panel shows, for $N=200$ that the value of $\tau_{ij}$ where the transition seems to happen depends also on the initial conditions imposed on $p_i$. When the agents start with moderate opinions, closer to 0.5, it is easier for agreement to happen. However, if the agents start with an initial very strong opinion towards one of the options, agreement is only possible if trust starts also high, so that agents with different choices can still influence each other. The same basic behavior was observed in every case run in the tests of the trust model, showing those results are quite robust. While the CODA model would always lead to agreement in a complete graph and to polarization if interaction were local, with self-reinforcing neighborhoods, here we can observe both results in the same model,  depending on the initial level of trust in the population.

\begin{figure}
\centering
\begin{tabular}{cc}
\epsfig{file=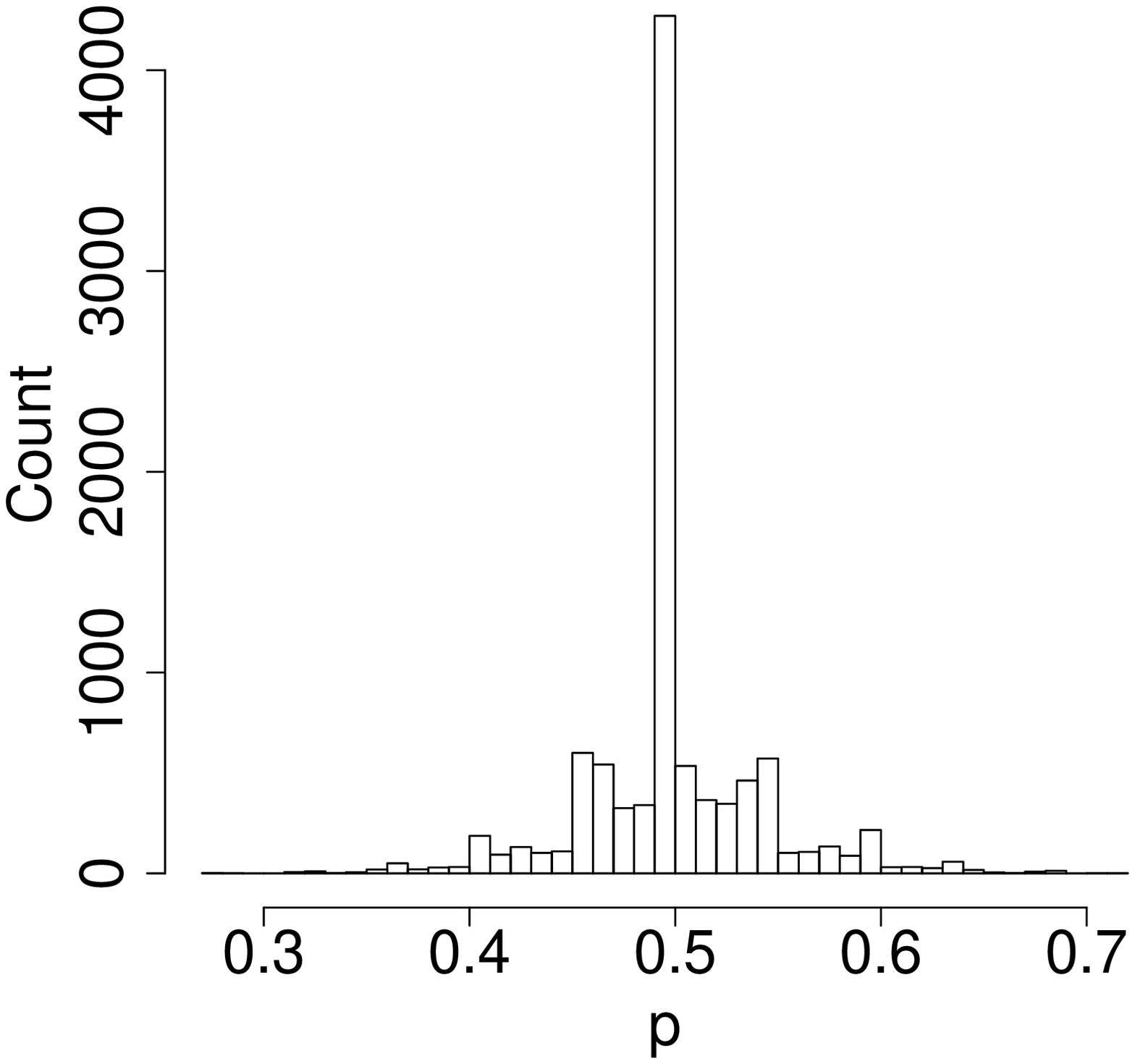,width=0.35\linewidth,clip=} & 
\epsfig{file=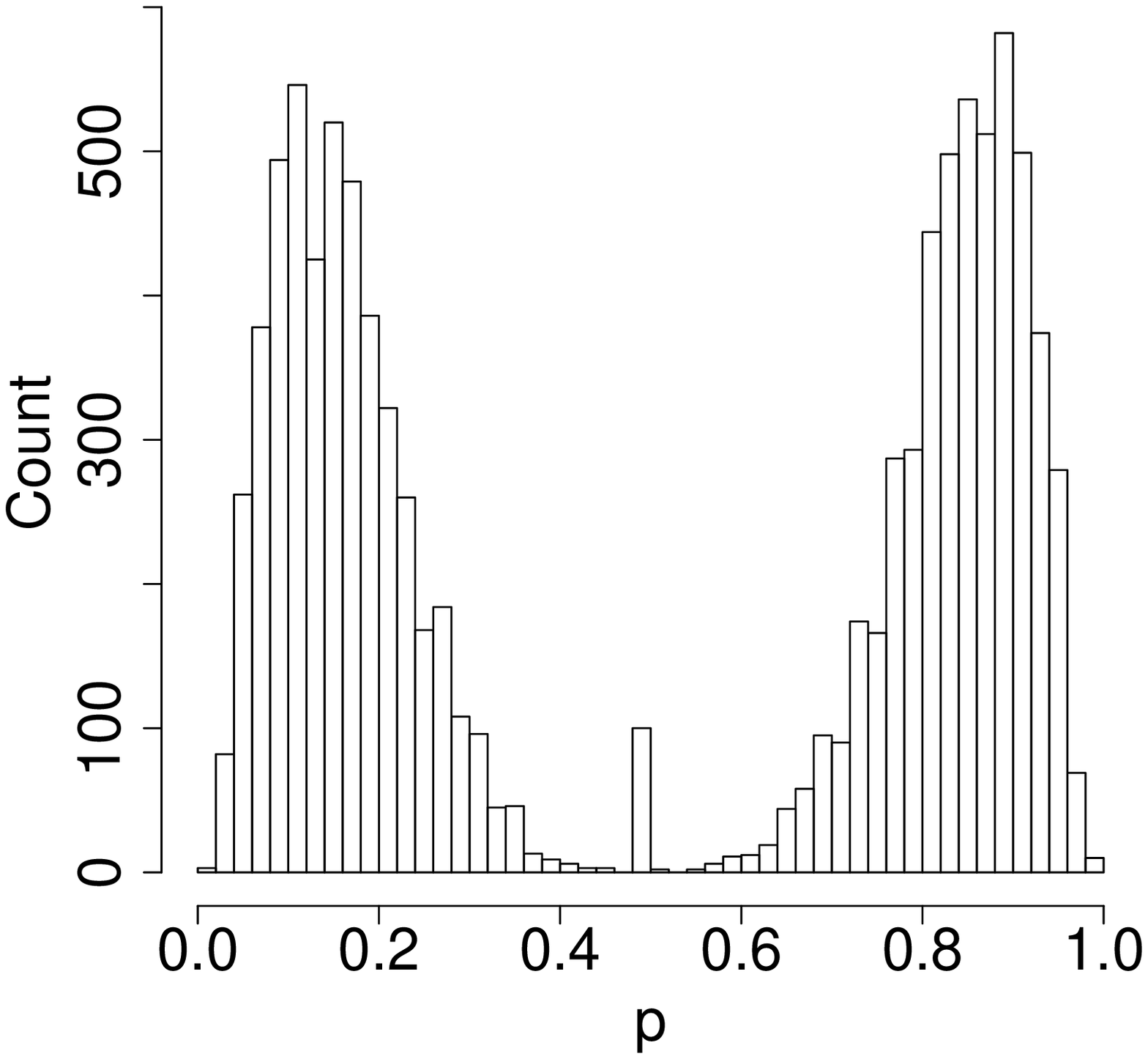,width=0.35\linewidth,clip=}
\end{tabular}
\caption{Temporal evolution of the histograms showing the distribution of observed $\tau_{ij}$ after different amounts of simulation time for $N=100$. In all cases, $\alpha=1-\mu=0.55$ and all agents started with $\tau_{ij}=0.5$ for all values of $i$ and $j$. {\it Left Panel}: $T=100$. {\it Right Panel}: $T=1,000$.}\label{fig:tau}
\end{figure}

The effects on the trust between the agents, on the other hand, have a much slower dynamics when compared to the dynamics of $p_i$. The reason for that is trivial. While each time an agent observes another, it updates its value of $p_i$, it updates just one of its $N$ values of $\tau_{ij}$. Therefore, trust evolves very slowly, as can be seen in Figure \ref{fig:tau} for a realization with only $N=100$ agents. It is clear that, even after $100$ average interactions per agent ($T=100$), the system is still very close to its initial condition (every $t_{ij}=0.5$). This happens because, in average, each value of the trust matrix $\tau_{ij}$ had only one update. This causes the high central peak in the left panel figure, corresponding to the large number of positions of $\tau_{ij}$ that were not updated at all. After $T=1,000$, that peak is much smaller and most agents have some evaluation about how much to trust most of others, trusting those who agree with them and distrusting the others.

Although the introduction of trust allows the existence of the two phases, the dynamics associated to $\tau_{ij}$ happens in a much larger time scale than the dynamics of $p_i$. That is, the decision if the system will converge to agreement or polarization happens before there was time enough for the trust matrix to change much. First, agents reinforce their opinions in one direction, simply by randomly been exposed to one choice more than the other. Once they start making up their minds, then they start trusting more those they agree with and less those they disagree.

The system as a whole seems to be dominated by two opposing forces, associated with different random walks. On one side, there is the tendency of the average of $p_i$ to change, causing a stronger influence towards one of the sides that would eventually lead to an agreement. On the other side, we have the tendency of individual values of $p_i$ to move away from the moderate region. And, as they get close to 0 or 1, agents tend to mistrust those who disagree with them and not change their opinions much when faced with disagreement. If the system remains in an undecided state long enough for the more extreme opinions to become important, polarization ensues. How much agents trust each other at first, therefore, determines which of these two opposing forces will win eventually.

\section{Conclusion}

We have seen that, by introducing trust in the CODA model, we obtain the realistic case where either agreement or polarization may be observed in the long run, depending on the parameters and initial conditions of the system. As it would be expected, the model says that when people trust each other more, agreement is easier to be obtained. Also the quality of the information plays an important role, in the sense of which $\alpha$ or $\mu$ is closer to 0.5. If trustworthy agents information is considered less random, agreement is favored, otherwise, polarization is.

Finally it is interesting to stress the different time scales involved with the two processes. It is a prediction of this model and, as such, a way to test it that, when agreement will happen, it happens comparatively fast. On the other hand, if the state of doubt remains for very long time in the society, this allows mistrust to build, eventually leading to polarization.

\vspace{1.0cm}
{\it Acknowledgments}

The author would like to thank the Santa Fe Institute for providing the facilities where the ideas presented here were finalized and the first simulations of the problem were performed.

\bibliographystyle{unsrt}
\bibliography{biblio}

\end{document}